\documentclass[twocolumn,secnumarabic,amssymb,nobibnotes,aps,prl,showpacs,superscriptaddress]{revtex4-1}
\usepackage[english]{babel}
\usepackage{graphicx}
\usepackage{amsmath}
\usepackage{amsfonts}
\usepackage[usenames,dvipsnames]{color}
\usepackage{hyperref}
\usepackage{blindtext}
\hypersetup{colorlinks}
\hypersetup{linkcolor=black, citecolor=black, urlcolor=blue}
\usepackage[utf8]{inputenc}
\usepackage{color}

\begin{document}
\title{Topological-defect-induced surface charge heterogeneities in nematic electrolytes}
\author{Miha Ravnik}
\address{Faculty of Mathematics and Physics, University of Ljubljana, Jadranska 19, 1000 Ljubljana, Slovenia}
\address{Department of Condensed Matter Physics, Jozef Stefan Institute, Jamova 39, 1000
Ljubljana, Slovenia}
\author{Jeffrey C. Everts}
\email{jeffrey.everts@gmail.com}

\address{Faculty of Mathematics and Physics, University of Ljubljana, Jadranska 19, 1000 Ljubljana, Slovenia}

\date{\today}

\begin{abstract}
We show that topological defects in an ion-doped nematic liquid crystal can be used to manipulate the surface charge distribution on chemically homogeneous, charge-regulating external surfaces, using a minimal theoretical model. In particular, the location and type of the defect encodes the precise distribution of surface charges and the effect is enhanced when the liquid crystal is flexoelectric. We demonstrate the principle for patterned surfaces and charged colloidal spheres. More generally, our results indicate an interesting approach to control surface charges on external surfaces without changing the surface chemistry.
\end{abstract}

\maketitle
The precise distribution of bound electric charges on a surface, space curve, or a (crystal) lattice, has important consequences in diverse fields ranging from biology, chemistry, and material science. On the (macro)molecular scale, differences in electronegativity of atoms \cite{Pauling:1932} result in a heterogeneous electron distribution in molecules which affects the acidity and, therefore, chemical reactivity of (organic) functional groups \cite{Fox:2004}. 
Surface charge distributions in biological systems can affect protein-protein and protein-ligand interactions \cite{Honig:1995, Paulini:2005, Li:2015} and virus assembly \cite{Siber:2012, Zandi:2017}. Similarly, for colloidal particles, heterogeneous surface charge distributions can  affect pair interactions \cite{Chen:2009, Levin:2019}, their self-assembly \cite{Bianchi:2014, Sabapathy:2017, Cruz:2016, Everts:2016b}, and catalytic properties \cite{Huang:2015} with direct relevance in applications such as nanoparticle-based drug delivery \cite{Salonen:2012}. On even larger length scales, surface charge heterogeneities on flat plates can cause spatial ion-concentration oscillations \cite{Gillespie:2017}, influence electrokinetic flow \cite{Ajdari:1995, Xie:2020}, and affect device performance \cite{Zhang:2009} in polycrystalline batteries \cite{Xu:2020} or solar cells \cite{Ni:2020}. In some materials charge modulations can even induce superconductivity \cite{Guinea:2018, Pelc:2019} or colossal magnetoresistance \cite{Miao:2020}. Overall, as clearly evident from these examples, the control and manipulation of charge profiles is highly relevant and an open challenge in diverse fields of science and technology.
 
Surface charge heterogeneities of ionic or electronic nature, can be realised in various ways. In colloidal science, a fixed surface charge distribution can be realised with different-surface charging functionalities on the same particle, such as patchy particles \cite{Glotzer:2007, Bianchi:2017} or Janus particles \cite{Hong:2006}, whereas charge-heterogeneous flat surfaces can be manufactured with agents such as micelles \cite{Klein:2012}. These type of charge heterogeneities are effectively permanent and imprinted in the material properties, but they can also be induced by ion packing \cite{Li:2012}, dielectric contrast \cite{Luijten:2014, Everts:2016}, many-body effects \cite{Everts:2016b}, particle shape \cite{Everts:2018} or flow \cite{Werkhoven:2018}. In all these cases the control over the surface charge distribution is, however, limited.

\begin{figure}[t]
\centering
\includegraphics[width=0.5\textwidth]{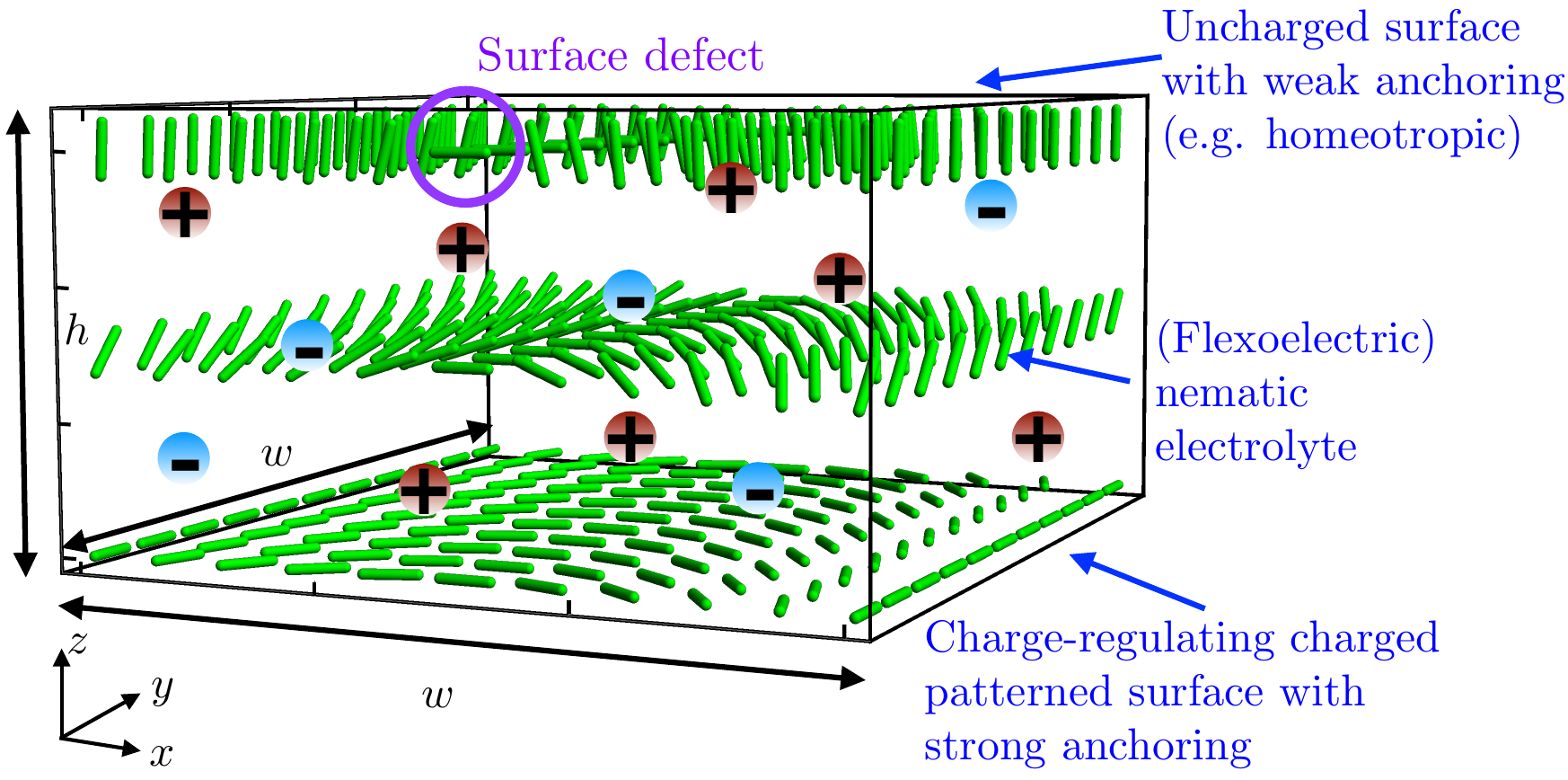}
\caption{Scheme of a nematic cell with a patterned, charge-regulating bottom plate and an uncharged top plate. The cell is filled with a (non)flexoelectric nematic electrolyte. We show an example of a pattern discussed in the main text, and for selective values of $z$ the associated director pattern as green rods obtained from numerical calcularions. In this example, the top plate exhibits a surface (boojum) defect.}
\label{fig:scheme}
\end{figure}

\begin{figure*}[t]
\centering
\includegraphics[width=\textwidth]{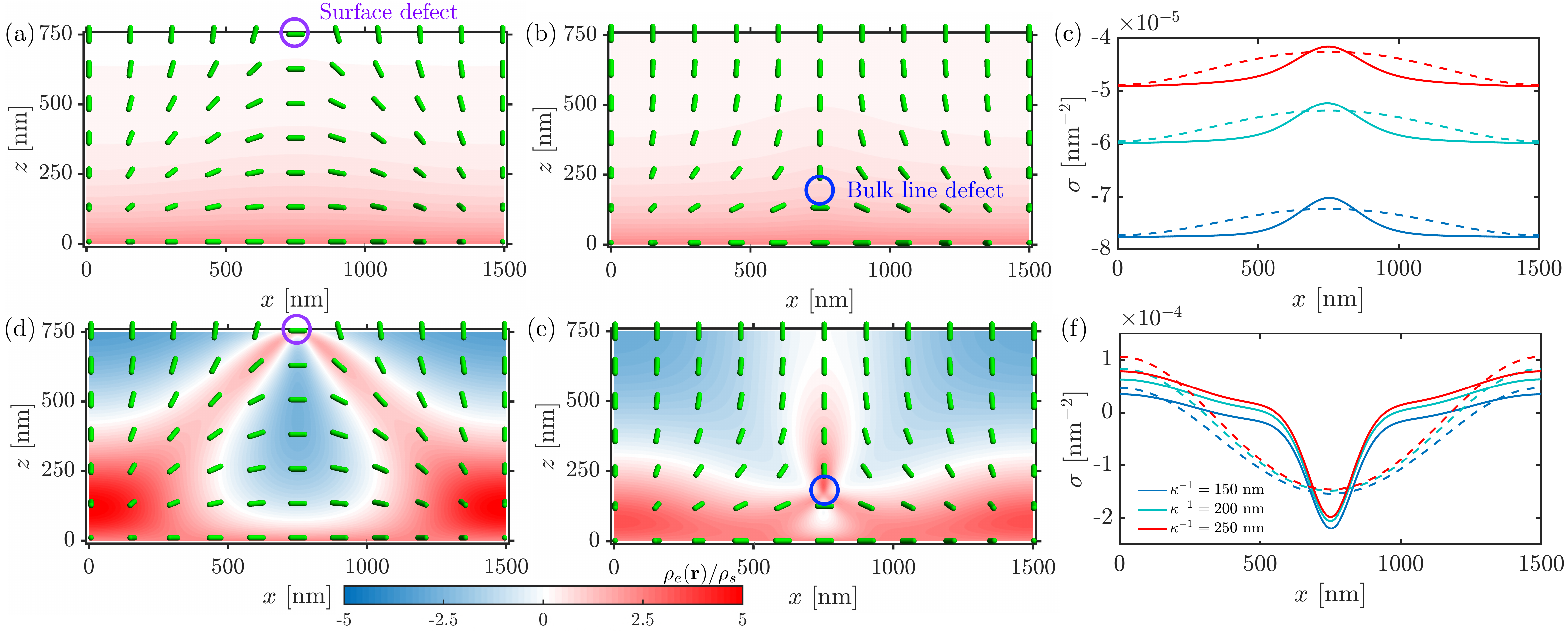}
\caption{Topological-defect induced surface charge heterogeneity on the bottom plate with constant surface potential $\Phi_0=-1$, in case of (a)-(c) no flexoelectricity $G=0$ and (d)-(f) $q_eG=10\ \mathrm{pC}\ \mathrm{m}^{-1}$. The top plate has homeotropic anchoring with a surface defect for $W=10^{-3}\ \mathrm{J}\ \mathrm{m}^{-2}$ [(a),(d)] or a bulk $-1/2$ line defect for $W=10^{-2}\ \mathrm{J}\ \mathrm{m}^{-2}$ [(b),(e)], which results in inhomogeneous diffuse net ionic charge densities $\rho_e({\bf r})$ (colormap). The proximity and type of defect result also in an inhomogeneous surface charge $q_e\sigma$ [(c),(f)], where dashed lines indicate the surface-defect case and full lines the bulk line defect case.}
\label{fig:cplinedefect}
\end{figure*}

In this Letter, we show that topological defects in ion-doped nematic fluids can be used to manipulate and spatially control the surface charge density of charge-regulating external surfaces, using a minimal numerical model. The position of topological defects in nematic fluids can be controlled by an external field, geometry and/or material flow, and the defects can be of diverse forms, from points, loops, networks to even knots and solitons \cite{Lopez:2011, Cates:2011, Alexander:2012, Nikkhou:2015, Mundoor:2016, Stebe:2016, Ackerman:2017, Rahimi:2017, Li:2018, Jones:2019, Mundoor:2019}, which we use in selected geometries. As the first geometry, we explore nematic cells consisting of parallel flat plates with a homogeneous charging functionality, but with an inhomogeneous surface (anchoring) imposed director pattern that generates surface or bulk topological defects. Today, such patterned surfaces can be experimentally realised with techniques such as photolitography \cite{Rastegar:2001, Kim:2018} or metasurfaces \cite{Guo:2017}. Secondly, we consider charged spherical colloidal particles in nematic electrolytes, and show that the surrounding topological defect breaks the spherical symmetry of the surface charge causing a surface charge inhomogeneity. In both examples, we also show that the topological-defect induced charge heterogeneity is further enhanced if the nematic is flexoelectric, meaning that a local elastic distortion in the orientational order of the nematic fluid causes a local electric polarisation \cite{Meyer:1969, Harden:2006, Castles:2010, Copic:2018}. More generally, this work shows that by designing and manipulating the surface or bulk topological defects in nematic electrolytes, one could realise and controllably tune diverse -possibly arbitrary- surface charge profiles.

An inhomogeneous surface charge profile is first demonstrated in the geometry of a standard nematic cell as shown in Fig. \ref{fig:scheme}, which consists of two parallel flat plates, but each with different surface anchoring: the top surface is assumed to impose homogeneous homeotropic anchoring, whereas the bottom surface is set to impose the indicated patterned anchoring. The cell is filled with a nematic electrolyte, characterised by tensorial order parameter ${\bf Q}({\bf r})$ \cite{deGennes:1993}, and positive and negative {\color{black} monovalent} ions with densities $\rho_{\pm}({\bf r})$. The top plate $(z=h)$ is uncharged, whereas the bottom plate ($z=0$) is charged, with surface charge density $q_e\sigma({\bf r})$, and $q_e$ the proton charge. The top and bottom surface impose an inhomogeneous nematic (director) profile -- both at the surface and in the nematic bulk, that even includes a surface or a bulk defect --,  which causes the development of an inhomogeneous electrostatic potential $\phi({\bf r})/(\beta q_e)$, with $\beta^{-1}=k_BT$ the thermal energy. The ion distributions are given within the mean-field approximation by the Boltzmann distributions $\rho_\pm({\bf r})=\rho_s\exp[\mp\phi({\bf r})]$, with $\rho_s$ as the reservoir salt concentration, to write the modified Poisson-Boltzmann equation as \cite{SI},
\begin{align}
\nabla\cdot&\left[\left(\mathbf{I}+\frac{2}{3}\frac{\epsilon_a^m}{\bar{\epsilon}}{\bf Q}\right)\cdot\nabla\phi-4\pi\lambda_B{\bf P}_f\right]= \kappa^2\sinh\phi,
\label{eq:wedgepoisson}
\end{align}
with ${\bf I}$ the unit tensor, $\epsilon_a^m$ the molecular dielectric anisotropy, $\bar{\epsilon}$ the rotationally averaged dielectric tensor, $\lambda_B$ the isotropic Bjerrum length, and we used the single-constant approximation for the flexoelectric (and order electric) polarisation $q_e{\bf P}_f=q_eG\nabla\cdot{\bf Q}$, with $q_eG$ the molecular flexoelectric constant. Finally, $\kappa^{-1}=(8\pi\lambda_B\rho_s)^{-1/2}$ is the isotropic (reference) Debye screening length. We use typical values of thermotropic nematics $\bar{\epsilon}\sim10$ \cite{Bogi:2001}, with $\kappa^{-1}$ between $10^2$ and $10^3$ nm (bulk ion densities of $10^{-8}-10^{-6}$ M) \cite{Thurston:1984b, Shah:2001, Musevic:2002, Everts:2020p}, where the nonelectrostatic contribution of the ions can be approximated to behave as an ideal gas \cite{Valeriani:2010}. 

The electrostatics is fully coupled to the nematic $\bf Q$ tensor profile given by the minimum of the total free energy, where the balance between nematic elasticity, flexoelectricity, and dielectric anisotropy gives the following equation for ${\bf Q}$ \cite{SI}:
\begin{align}
\beta  L\nabla^2{\bf Q}+&G\overline{\nabla\otimes\nabla\phi}+\frac{\epsilon_m^a}{12\pi\lambda_B\bar{\epsilon}}\overline{\nabla\phi\otimes\nabla\phi}= \nonumber \\
&\left[\beta A+\beta C\mathrm{tr}({\bf Q}^2)\right]{\bf Q}+\beta B \overline{{\bf Q}^2},
\label{eq:wedgeQ}
\end{align}
with $L$ the nematic elastic constant (within the single-constant approximation) and Landau-de Gennes bulk parameters $A$, $B$, and $C$. For all these parameters, we take the values corresponding to standard nematic liquid crystals, like 5CB \cite{SI}, unless stated otherwise. The overline defines the traceless part of the tensor, e.g., $\overline{\bf A}={\bf A}-(\mathrm{tr}{\bf A}){\bf I}/3$ for an arbitrary $3\times3$ tensor ${\bf A}$ \footnote{In this letter we solve Eqs. \eqref{eq:wedgepoisson}-\eqref{eq:wedgeQ} \emph{simultaneously}; however, for the geometries with relatively strong anchoring in this letter, the second and third term in Eq. \eqref{eq:wedgeQ} are small compared to the elastic and bulk contributions, but are important at higher voltages \cite{deGennes:1993} or sufficiently weak anchoring \cite{Onuki:2009, Everts:2020p}.}. Eqs. \eqref{eq:wedgepoisson} and \eqref{eq:wedgeQ} are akin to models of dilute electrolytes in binary fluid-fluid mixtures with ions coupled to the fluid composition profile \cite{Onuki:2006, Samin:2012, Bier:2012, Everts:2016}. Note, that in our approach we generalise and solve this model for nematic electrolytes with the full nematic tensorial order parameter, dielectric anisotropy, and flexoelectricity included.

The charge at the bottom plate is caused by the ad- or desorption of ions \cite{Ninham:1971} from the nematic fluid on specific chemical groups located on this surface, where we assume the bottom plate to be chemically homogeneous in the sense that these chemical groups are identical over the entire surface and homogeneously distributed. Now the local surface charge depends on the local ion concentration; therefore, any inhomogeneity in the ion concentration will lead to an inhomogeneous surface charge.  We will show that such inhomogeneities can be realised by an inhomogeneous nematic ${\bf Q}$-tensor profile. We model the bottom plate as a constant-potential plate with boundary condition $\phi(x,y,0)=\Phi_0$, being the simplest form of a charge-regulating electrostatic boundary condition, with $\Phi_0$ as the surface-imposed constant potential. The surface charge profile $\sigma$ can then be evaluated by the normal component of the dielectric displacement,
\begin{align}
\sigma=-\hat{\boldsymbol{\nu}}\cdot\Bigg[\left({\bf I}+\frac{2}{3}\frac{\epsilon_a^m}{\bar{\epsilon}}{\bf Q}\right)\cdot\nabla\phi/(4\pi\lambda_B)-G\nabla\cdot{\bf Q}\Bigg],
\label{eq:surfcharge}
\end{align}
with $\hat{\boldsymbol{\nu}}$ as an outward pointing unit normal vector, and assuming no electric fields are generated inside the bottom plate.

Although, we assume that chargeable chemical groups are homogeneously distributed, the bottom plate is patterned in terms of the nematic anchoring, imposing a fixed orientational director pattern ${\bf n}^B$. We assume strong anchoring conditions for the bottom plate, and set ${\bf Q}={\bf Q}^B$with $S_\text{eq}$ the equilibrium (bulk) value of the scalar order parameter, and no surface-imposed biaxiality. Specifically, we consider the one-dimensional pattern 
 \begin{equation}
 {\bf n}^B=(\sin(x\pi/w),\cos(x\pi/w),0),
 \label{eq:I}
 \end{equation}
periodic in $x$ with period $w$, and a top plate with homeotropic anchoring strength $W$ \cite{Durand:1992}. Note that we consider periodic surface imposed anchoring patterns, but -in view of the mechanism for the creation of surface charge heterogeneities- they could also be realised with nonperiodic and even random patterns.  

%

Fig. \ref{fig:cplinedefect} shows the topological-defect induced surface charge heterogeneity as generated by the introduced surface anchoring pattern [Eq. \eqref{eq:I}] obtained from numerical calculations with the finite-element software package COMSOL Multiphysics. For smaller anchoring [Fig. \ref{fig:cplinedefect}(a) and Fig. \ref{fig:scheme}], the nematic structure exhibits a a surface defect line, whereas for stronger anchoring [Fig. \ref{fig:cplinedefect}(b)], the defect evolves into a -1/2 defect line along the $y$ direction. For no flexoelectricity ($G=0$), the diffuse ionic screening cloud $\rho_e({\bf r})=\rho_+({\bf r})-\rho_-({\bf r})$ is translational invariant in $y$ but not in $x$ and $z$, and is more extended at fixed $\kappa^{-1}$ in the case of the bulk line defect. However, the most striking observation is that not only the screening cloud is affected by the location of the defect, but also the surface charge density, see Fig. \ref{fig:cplinedefect}(c), which becomes strongly peaked around the location of the defect when the defect approaches the bottom plate; compare dashed (surface defect) with full (bulk defect) lines, and this behaviour is robust also for varying screening length $\kappa^{-1}$. This result clearly highlights, that surface charges can be induced by manipulating topological defects, such as by bringing a nematic defect closer to the surface. The overall increase in $\sigma$ upon decreasing $\kappa^{-1}$ can be understood from the Grahame equation \cite{Grahame:1947}, $\sigma=\sinh\left(\Phi_0/2\right)/(2\pi\lambda_B\kappa^{-1})$,
which is an exact result for isotropic solvents and infinite-plate separation, but turns out to be a reasonable estimate in calculating the average surface charge density even for the nematic case \cite{SI}.

\begin{figure}[t]
\centering
\includegraphics[width=0.5\textwidth]{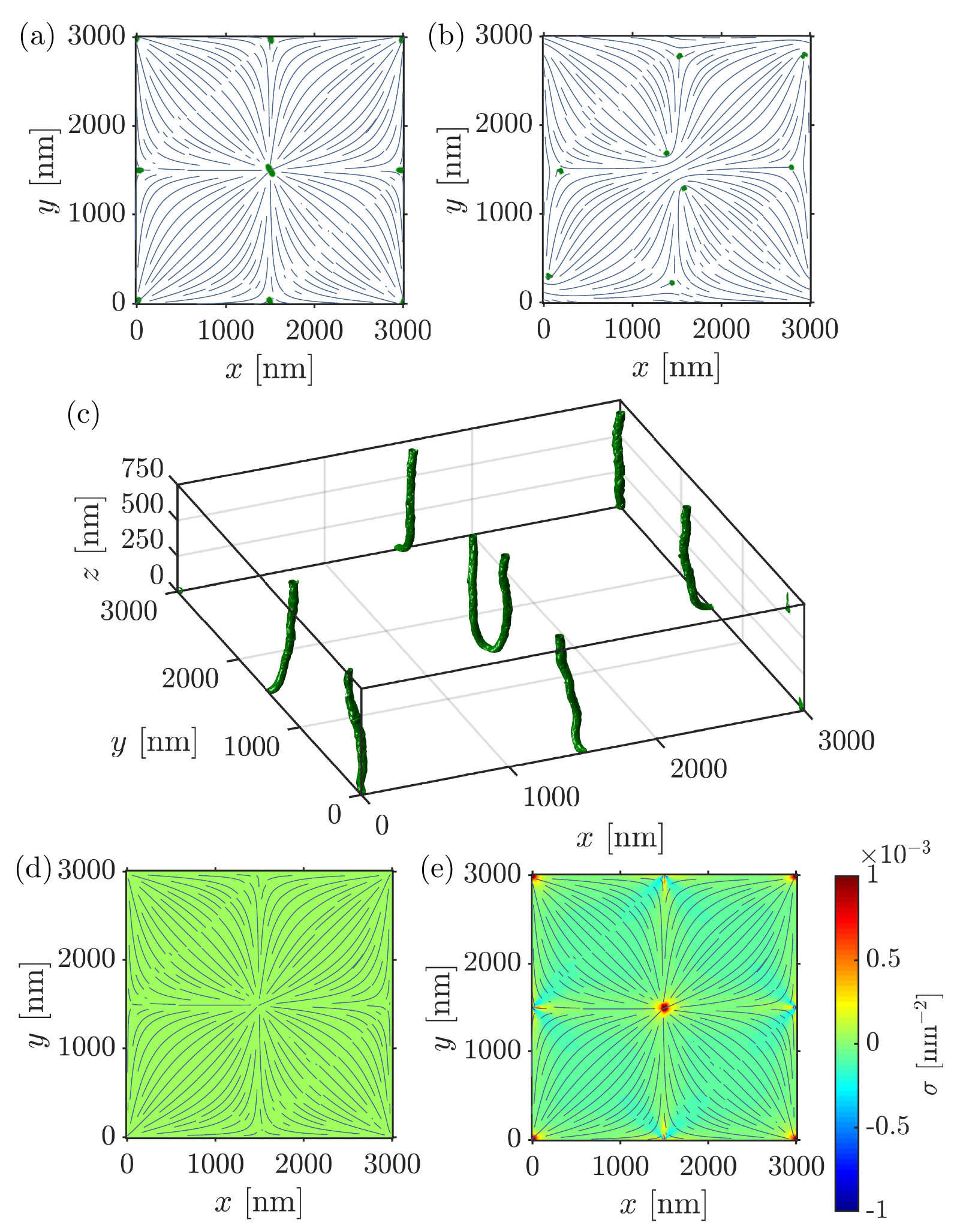}
\caption{Two-dimensional in-plane nematic distortion for the generation of surface charge heterogeneities. Topological defects are shown in green ($S<0.48$) for (a) close to bottom surface at $z=10$ nm and (b) top surface at $z=750$ nm. (c) Three-dimensional isosurfaces of $S=0.48$ indicate the location of the line defects. Structure is for parameters $w=1500$ nm and $h=750$ nm. The top plate has weak planar degenerate anchoring conditions with $W_1^d=W_2^d=10^{-5}\ \mathrm{J} \ \mathrm{m}^{-2}$. (d,e) Surface charge distributions at $\kappa^{-1}=250$ nm and fixed surface potential $\Phi_0=-1$ with (d) the nematic being not flexoelectric ($G=0$) and in (e) with $q_eG=10\ \mathrm{pC \ m}^{-1}$.}
\label{fig:structureII}
\end{figure}

Flexoelectricicty in a nematic electrolyte has a further major effect on the diffuse ion cloud \cite{Everts:2020} and in turn on the surface charge heterogeneity. Figs. \ref{fig:cplinedefect}(d-e) show that the surface charge profile at the bottom plate $\sigma$ becomes more inhomogeneous as the result of flexoelectricity  [Fig. \ref{fig:cplinedefect}(f)], and bringing the defect closer to the bottom surface results in a local minimum in $\sigma$ rather than a local maximum [compare to Fig. \ref{fig:cplinedefect}(c)]. The trend with varying $\kappa^{-1}$ is similar as for $G=0$ for the average surface charge density. The manipulation of surface charges by topological defects is robust with respect to the type of charge-regulating boundary condition, as charge localisation around topological defects is observed also if we assume an ion-dissociation boundary condition on the bottom plate, see \cite{SI}.

To generalise, the key requirement for realising surface charge heterogeneities, is that the director component perpendicular to the surface has a gradient which varies across the surface. We demonstrate this requirement in a cell with a patterned bottom plate of the form 
\begin{equation}
{\bf n}^B=\dfrac{(\sin(x\pi/w),\sin(y\pi/w),0)}{\sqrt{\sin^2(x\pi/w)+\sin^2(y\pi/w)}},
\label{eq:II}
\end{equation}
and planar degenerate anchoring at the top cell surface $z=h$, with $W_1^d$ the in-plane anchoring strength, and $W_2^d$ the surface-ordering anchoring strength \cite{Fournier:2005}. Such a nematic cell has a strictly inplane $(xy)$ director field ($n_z=0$), see Figs. \ref{fig:structureII}(a)-(b) for bottom and top plate profiles. In Fig. \ref{fig:structureII}(c), we show the three-dimensional structure of the nematic defect lines, with +1 defects in the corners and in the centre of the unit cell, and  $-1$ defects in the center of each edge of the cell for $z=0$. The $\pm1$ defects split further away from the bottom surface into two $\pm1/2$ defects. For no flexoelectricity $G=0$, we find a roughly homogeneous surface charge distribution [Fig. \ref{fig:structureII}(d))], whereas for the flexoelectric case $G\neq0$ gradients in the electrostatic potential and gradients in $S$ lead to an inhomogeneous surface charge distribution, with positive peaks around the defects. We find that the absence of director field gradients perpendicular to the surface leads to much weaker -but now twodimensional- control of the surface charge by topological defects, in contrast to the one-dimenensional pattern of Eq. \eqref{eq:I}.
\begin{figure}[t]
\centering
\includegraphics[width=0.49\textwidth]{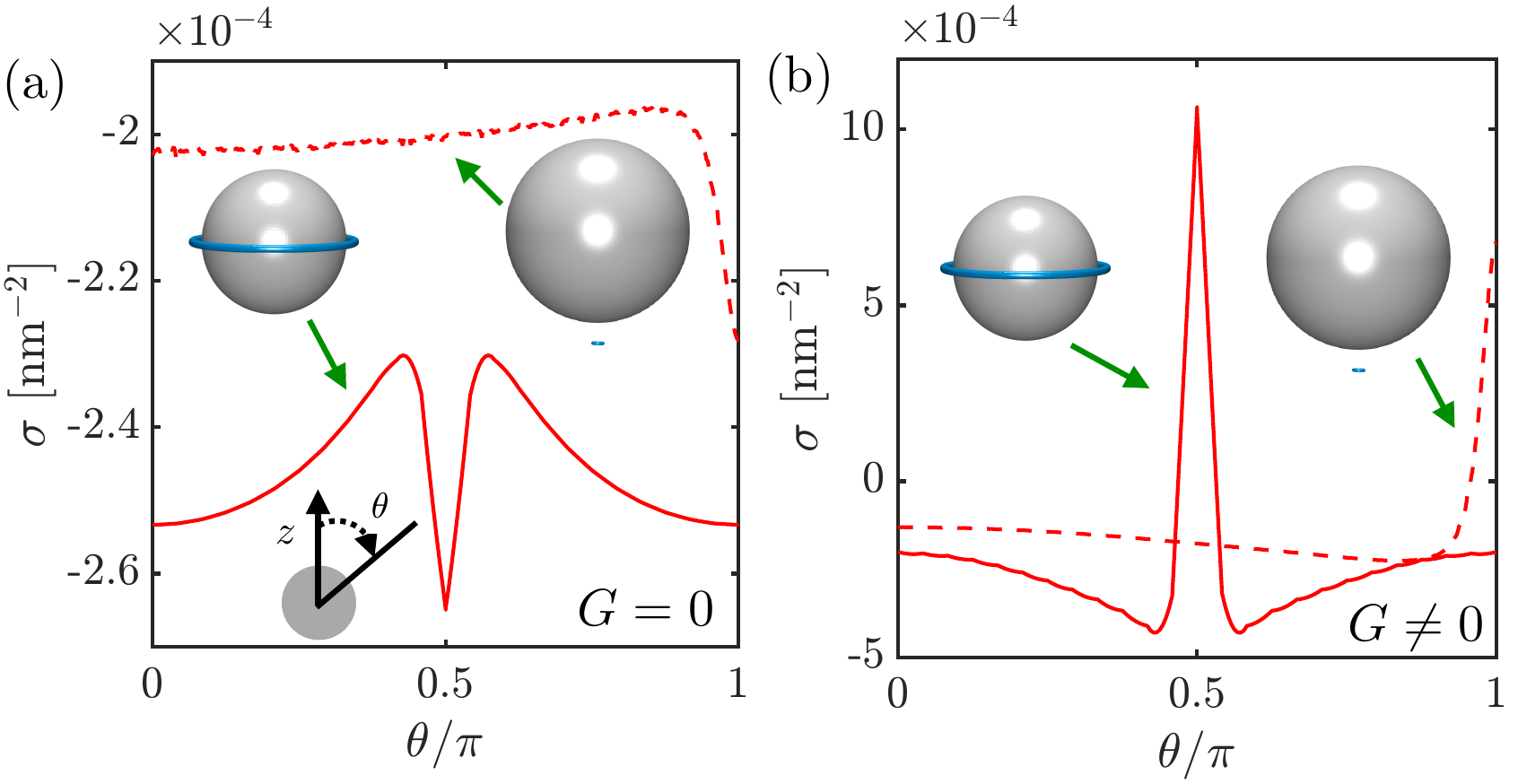}
\caption{Heterogeneous surface charge distributions $q_e\sigma$ of a constant-potential colloidal sphere ($\Phi_0=-1$) with dielectric constant $\epsilon_p=2$ and strong anchoring conditions in a nematic electrolyte with $\kappa^{-1}=100$ nm. (a) For $G=0$, we show the surface charge profile in the case of the nematic Saturn ring defect (full line, particle radius $r_p=250$ nm) or the point defect (dashed line, particle radius $r_p=1000$ nm). In (b) we show the enhancement of the surface charge heterogeneity by flexoelectricity with $q_eG=10 \ \mathrm{pC \ m^{-1}}$.}
\label{fig:sigmacolloid}
\end{figure}

As the third system, we show that even homogeneous charging functionalised surfaces with \emph{homogeneous anchoring} conditions can result in surface charge heterogeneities due to topological defects, emergent from the geometry and topology of the surface. The simplest system where this occurs is a charged colloidal sphere with homeotropic (perpendicular) anchoring, immersed in a homogeneous nematic with far-field director in the $z$ direction. Such charged spheres have only been theoretically investigated in the absence of ions and charge regulation \cite{Onuki:2009}. Depending on the particle size, a saturn-ring defect is formed (small particles) or a point defect (large particles), to compensate the distortion imposed by the particle, see the insets of Fig. \ref{fig:sigmacolloid} \cite{Abbott:2000, Stark:2001}. For particle radius $r_p=250$ nm, we see in Fig. \ref{fig:sigmacolloid}(a) that an inhomogeneous surface charge $\sigma$ is induced with two maxima and a minimum exactly at the equator for $G=0$ for the Saturn-ring defect, with an amplitude that is relatively small compared to the average. For nonzero flexoelectricity $G\neq0$, the amplitudes of the surface charge are enhanced and a maximum positive charge density is found at the equator (although $\Phi_0<0$). For the point defect, (e.g. $r_p=1\ \mu\mathrm{m}$), we find similar to the Saturn-ring defect a minimum of relatively small amplitude close to the location of the defect. This amplitude is enhanced for $G>0$ and the minimum turns into a maximum. These results show that particles of diverse shapes and topologies in nematic electrolytes could possibly realise diverse and highly heterogeneity-rich surface charge profiles. 

The major experimental challenge for exploring the defect-induced surface charge heterogeneities, is how to directly measure the surface charge distribution, rather than indirectly by e.g. considering the effective pair potential (colloidal particles) or disjoining pressure (flat plates). Most surface charge measurements rely on measuring the total surface charge, rather than the distribution, such as with titration \cite{Yates:1980} or electrophoresis \cite{Vissers:2011}, although there are a few cases where heterogeneities could still be assessed with these methods \cite{Schiller:1980, Gelfi:1985, Miklos:2004}. Other methods rely on flow \cite{Lis:2014, Werkhoven:2018}, or atomic-force microscopy \cite{Yin:2008}, but interpreting the results in these nematic systems might require additional mathematical modelling \cite{Hartkamp:2018}. Furthermore, fluorescence-based techniques might offer a versatile route to locally probe the surface charge \cite{Chow:1980, Robinson:2013}. Finally, we mention surface plasmon resonance imaging, which is a promising probe-free spectroscopic technique that can map out the electric double layer \cite{Chen:2017, Luo:2018}.

On the theoretical side we neglected effects that become more prevalent at high ion densities and/or lower $\bar{\epsilon}$, such as Bjerrum pairs \cite{Levin:1993, Valeriani:2010, Adar:2017}, steric effects \cite{Bikerman:1942, Orland:1997, Onuki:2017}, and ion-specific interactions \cite{Levin:2009}, that are all well investigated for isotropic solvents but not for nematic electrolytes. Furthermore, explicit ion dependence of the dielectric tensor, flexoelectric coefficients, and the liquid-crystal bulk phase behaviour are all relevant effects that deserve further study, as well as higher-order dipolar corrections to the Poisson-Boltzmann equation \cite{Orland:2007}. Furthermore, the presented results are relevant for lyotropic systems, which also exhibit dielectric anisotropy \cite{Straley:1973} and flexoelectricity \cite{Petrov:2006}. Likely, lower Debye lengths in lyotropics could be reached than in thermotropics because they are often water based and, therefore, have a higher ion solvability. Lyotropic building blocks can be colloidal particles of various designed shape and charge, therefore we envisage that lyotropics can have ion-tunable flexoelectricity and dielectric anisotropy, similar to the ion- and charge-tuning of the elastic constants of charged hard rods \cite{Drwenski:2016}.

In conclusion, our results show that topological defects in nematic electrolytes can be used to control and manipulate surface charges in a broad and extensive way, which can be further enhanced by the effects of flexoelectricity. Our work has clear relevance for the self-assembly of charged colloidal particles in liquid crystals \cite{Mundoor:2016}, but also on the performance of liquid-crystal-based devices, such as nematic-based electric double layer capacitors or nematic microfluidic applications \cite{Tkalec:2020}. Likely equally relevant, this work can be relevant for biological systems and active matter, where topological defects are known to have a prime role in their dynamic behaviour; in combination with charges this could lead to phenomena such as dynamic or active surface charge regulation. 

\begin{acknowledgments}
\emph{Acknowledgements} - J. C. E. acknowledges financial support from the European Union's Horizon 2020 programme under the Marie Skłodowska-Curie Grant Agreement No. 795377. M. R. acknowledges financial support from the Slovenian Research Agency ARRS under Contracts P1-0099, J1-1697, and L1-8135. The authors acknowledge fruitful discussions with S. \v Copar and B. Werkhoven.
\end{acknowledgments}
\bibliographystyle{apsrev4-1} 
\bibliography{literature1} 

\end{document}